\documentclass[10pt,twocolumn,letterpaper]{article}
\usepackage[iccvfinal]{template/iccv}      
\usepackage{url}
\usepackage{natbib}
\usepackage{float}

\definecolor{iccvblue}{rgb}{0.21,0.49,0.74}
\usepackage[pagebackref,breaklinks,colorlinks,allcolors=iccvblue]{hyperref}
\usepackage{comment}
\usepackage{multirow}
\def\paperID{11719} 
\def\confName{ICCV}
\def\confYear{2025}

\title{VoteSplat: Hough Voting Gaussian Splatting for 3D Scene Understanding}

\author{
Minchao Jiang$^{1*}$, Shunyu Jia$^{1*}$, Jiaming Gu$^{1,2}$, Xiaoyuan Lu$^{3}$, 
Guangming Zhu$^{1}$,\\ Anqi Dong$^{4}$, Liang Zhang$^{1\dagger}$ \\
$^{1}$ School of Computer Science and Technology, Xidian University\\
$^{2}$ Algorithm R\&D Center, Qing Yi (Shanghai)\\
$^{3}$ Shanghai Pudong Cryptography Research Institute\\
$^{4}$ Division of Decision and Control Systems and Department of Mathematics,\\ KTH Royal Institute of Technology\\
{\tt\small \{jamchaos, syjia\_2001\}@stu.xidian.edu.cn}, 
{\tt\small jiaming\_gu\_xidian@outlook.com}, \\
{\tt\small \{gmzhu, liangzhang\}@xidian.edu.cn},
{\tt\small xylu@bnc.org.cn}, {\tt\small anqid@kth.se}
}

\begin{document}
\twocolumn[{%
\renewcommand\twocolumn[1][]{#1}%
\maketitle
\begin{center}
\centering
\includegraphics[width=1\textwidth]{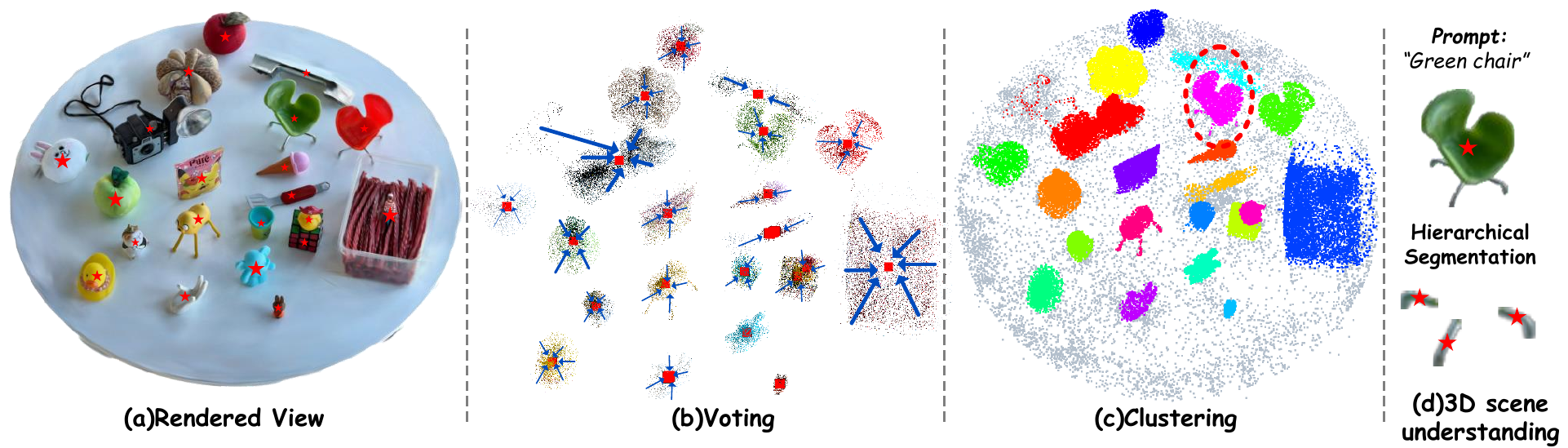}
\captionof{figure}{
VoteSplat integrates 3DGS and Hough Voting for 3D scene understanding: (a) 3DGS retains its original novel view synthesis capability, (b) each Gaussian primitive encodes an offset vector which votes the point cloud to the instance center, (c) 3D vote clustering enables instance segmentation, and (d) open-vocabulary 3D instance localization and click-based 3D object localization are demonstrated.
}
\end{center}%
}]

\begin{abstract}
\renewcommand{\thefootnote}{\fnsymbol{footnote}}\footnotetext[1]{Contribute equally.}\footnotetext[2]{Corresponding author.}3D Gaussian Splatting (3DGS) has become horsepower in high-quality, real-time rendering for novel view synthesis of 3D scenes. However, existing methods focus primarily on geometric and appearance modeling, lacking deeper scene understanding while also incurring high training costs that complicate the originally streamlined differentiable rendering pipeline. To this end, we propose VoteSplat, a novel 3D scene understanding framework that integrates Hough voting with 3DGS. Specifically, Segment Anything Model (SAM) is utilized for instance segmentation, extracting objects, and generating 2D vote maps. We then embed spatial offset vectors into Gaussian primitives. These offsets construct 3D spatial votes by associating them with 2D image votes, while depth distortion constraints refine localization along the depth axis. For open-vocabulary object localization, VoteSplat maps 2D image semantics to 3D point clouds via voting points, reducing training costs associated with high-dimensional CLIP features while preserving semantic unambiguity. Extensive experiments demonstrate VoteSplat’s effectiveness in open-vocabulary 3D instance localization, 3D point cloud understanding, click-based 3D object localization, hierarchical segmentation, and ablation studies. Our code is available at 
\href{https://sy-ja.github.io/votesplat/}{VoteSplat}.
\end{abstract}

\section{Introduction}\label{sec:intro}

Localization and instance-level semantic understanding of 3D scenes are critical objectives in the computer vision community, especially with the rise of embodied intelligence. Traditional point cloud-based methods for 3D scene understanding have been widely explored, including classification~\cite{qi2017pointnet}, segmentation~\cite{qi2017pointnet++}, and detection~\cite{qi2019deep}. Recently, 3D Gaussian Splatting (3DGS)~\cite{kerbl20233d} has gained popularity for its ability to achieve real-time, photorealistic novel view synthesis (NVS) at high resolutions. Unlike implicit Neural Radiance Fields (NeRF)~\cite{mildenhall2021nerf}, which represent scenes as continuous volumetric functions, 3DGS reconstructs scene appearance and geometry using point-clouds-alike Gaussian primitives. Building on its success, 3DGS has been extended to various domains, including rendering \cite{yu2024mip, lu2024scaffold,dong2024global}, surface reconstruction~\cite{huang20242d, chen2024pgsr, yu2024gaussian}, generation~\cite{yi2023gaussiandreamer}, and scene understanding~\cite{qin2024langsplat,zhai_cvpr25_panogs}. Moreover, scene reconstruction and semantic understanding can be seamlessly integrated with 3DGS rendering pipeline, facilitating more advanced intelligent agent interaction and decision-making.

Efforts have aimed to integrate learnable semantic information into 3DGS, enhancing the language-grounded capabilities of 3DGS. Current approaches can be broadly categorized into the following two types:

{\bf \noindent (i)} The first approach~\cite{qin2024langsplat, shi2024language,ye2024gaussian,qiu2024gls} directly embeds semantic vectors into Gaussian primitives, project semantic information onto images in the same way as rendering colors. This enables cross-frame association through semantics but has two key limitations~\cite{wu2024opengaussian}: (1) The high dimensionality of CLIP-extracted features leads to excessive training overhead, while dimensionality reduction introduces semantic ambiguity; (2) The object occlusion restricts them to pixel-level segmentation, where embedded semantic vectors are insufficient for point-level scene understanding.

{\bf \noindent (ii)} The second approach~\cite{wu2024opengaussian, choi2024click} adopts a point cloud clustering strategy to improve instance differentiation and reduce ambiguity, following a multi-stage pipeline. 3DGS independently reconstructs the scene and embeds feature vectors into the trained Gaussian primitives, using contrastive learning for instance differentiation. The goal is to address the spatial adjacency of primitives belonging to different instances, with additional features for better separation. However, contrastive learning adds significant computational complexity to the rendering pipeline.

To this end, we propose VoteSplat, that integrates Hough voting with Gaussian splatting for 3D scene understanding. VoteSplat defines a set of 3D Gaussians embedded with additional three-dimensional spatial offset vectors to compute spatial 3D votes. We expect 3D votes to be located at the centroid of the instances so that clustering methods can be directly applied to distinguish each instance without the need for additional learned features for differentiation. Given the Gaussian primitives forming a specific instance and its centroid are unknown, we propose 3D-2D Votes Association Learning as the centroid may exists in multi-view two-dimensional images. 

Specifically, for each image, we first apply the Segment Anything Model (SAM)~\cite{kirillov2023segment} to obtain well-segmented masks and compute their centroids with precise object boundaries. The pixel coordinates of these centroids serve as 2D ground-truth votes to supervise projection of 3D votes, encouraging convergence toward the instance center while ensuring multiview consistency. Since projection transformations cause depth information loss, relying solely on 2D vote supervision can introduce noise in spatial voting points. To overcome this, we introduce a depth distortion regularization term to improve spatial vote aggregation along the depth dimension.

In a trained VoteSplat scene, each Gaussian primitive surrounding an object has an offset vector pointing to a 3D vote near the instance centroid. Clustering these votes allows us to effectively determine instance IDs to the Gaussian primitives, which in turn establishes correspondences between point clouds and image semantics, enabling robust 3D scene understanding. The contributions can be thus summarized as 
\vspace{0.07in}
\begin{enumerate}
\item Hough voting is first considered into 3D Gaussian Splatting (3DGS) to achieve spatial clustering of point clouds belonging to the same instance. This enables point-level segmentation without requiring additional high-dimensional feature vectors, improving training efficiency and scene understanding accuracy.
\item 3D-2D Votes Association Learning is proposed, incorporating a custom depth distortion loss to enhance spatial aggregation of voting points along the depth dimension for denoising.
\item An instance ID-based approach is introduced to associate 2D image semantics with 3D point clouds, enabling the linking of CLIP features to individual 3D instances, and facilitating open-vocabulary scene understanding.
\end{enumerate} 
\vspace{0.07in}
Therefore, VoteSplat enables efficient point-level segmentation without requiring high-dimensional feature embeddings. Additionally, we introduce 3D-2D Votes Association Learning and depth distortion regularization to refine spatial clustering and improve localization accuracy. The rest of the paper is organized as follows: Section \ref{sec:relat} reviews related works in 3D Gaussian Splatting and Hough Voting. Section \ref{sec:metho} details our methodology, including 3D vote construction and semantic association. Section \ref{sec:exp} presents the experimental setup and results across various tasks. Finally, Section \ref{sec:con} concludes the paper.

\section{Related Works}
\label{sec:relat}
\paragraph{3D Gaussian Splatting for Scene Understanding.}
3D Gaussian Splatting  has emerged as a promising method for real-time scene rendering, offering superior visual quality.  Utilize 3DGS to jointly reconstruct the appearance and geometric information of a scene with instance and semantic information, to better support downstream tasks.

NeRF, as an innovative 3D reconstruction method, has inspired numerous works~\cite{siddiqui2023panoptic, zhi2021place} to develop 3D language fields upon it. LERF\cite{kerr2023lerf} first integrated CLIP\cite{caron2021emerging} features into NeRF, constructing language embedded radiance fields to enable open-vocabulary 3D querying. Additionally, DINO features were used to enhance boundary accuracy. However, due to high computational costs, NeRF-based methods face rendering performance bottlenecks, leading to the adoption of 3DGS~\cite{kerbl20233d} with semantic information. Building on 3DGS, LangSplat\cite{qin2024langsplat} employs an autoencoder to reduce CLIP feature dimensionality, embedding the compressed representations into Gaussian primitives, while incorporating a semantic hierarchy. \citet{shi2024language} use VQ-VAE to quantize high-dimensional CLIP features into discrete categories, converting semantic supervision into category-level supervision, thereby reducing computational overhead. Similarly,~\citet{shorinwa2024fast} constructs a 3D language field by mapping semantic features to discrete categories.

These methods achieve multi-view training through cross-frame semantic similarity. To reduce computational complexity, CLIP feature needs to be either dimensionally reduced or classified, which inevitably introduces semantic ambiguity. Other methods ~\cite{ye2024gaussian, wu2024opengaussian, choi2024click, li2024instancegaussian} rely on learnable features to distinguish instances and ultimately assign complete semantic information. These methods ensure semantic accuracy; however, intra-class and inter-class contrastive learning can be highly time-consuming.

\paragraph{Hough Voting for 3D Point Clouds.} The Hough transform (also, Hough voting), originates in late 1950s~\cite{hough1959machine}, convert the detection of simple patterns in point samples as peak detection in a parametric space. The Generalized Hough Transform~\cite{ballard1981generalizing} extends this concept to image patches, enabling the identification of complex objects. Hough voting has been widely applied in various tasks, including the implicit shape model\cite{leibe2008robust}, plane extraction from 3D point clouds\cite{borrmann20113d}, 6D pose estimation~\cite{sun2010depth}, and so forth.

Hough voting has been successfully integrated with advanced learning techniques. In 3D object detection, a common approach~\cite{woodford2014demisting, knopp2010orientation, velizhev2012implicit, knopp2011scene} is to adapt mature 2D detection algorithms, such as Faster R-CNN\cite{ren2016faster} or YOLO\cite{redmon2016you}, to 3D point clouds by generating proposals at each input point. However, a fundamental challenge arises: 3D sensors capture only surface data, meaning the object center often lies in empty space, far from the available points in the input point cloud. As a result, there are typically no input points near the object center, making it difficult for surface-based networks to extract meaningful contextual information, leading to inaccurate proposals. To overcome this, ~\citet{qi2019deep} introduced VoteNet, a Hough voting-based method. VoteNet first samples seed points from the input point cloud and votes for the target’s center, generating voting points near the object center. These voting points are then used to generate bounding box proposals, effectively addressing the issue of inaccurate proposals when the object center is distant from the surface points.

\begin{figure*}[h]
    \centering
    \includegraphics[width=1\textwidth]{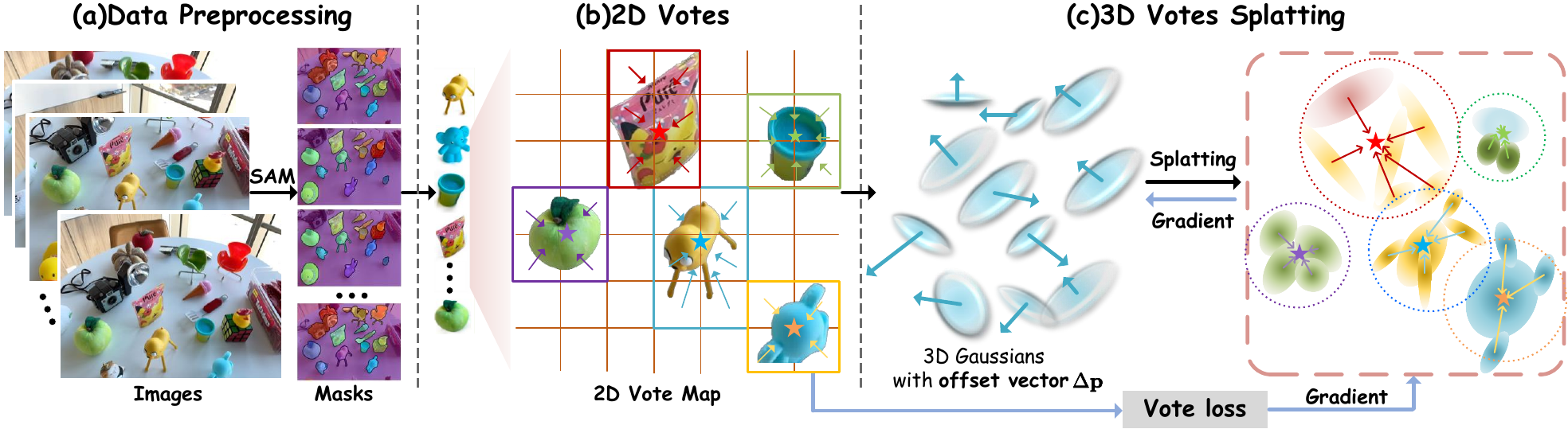} 
    \caption{Three main steps in VoteSplat pipeline: (a) We first deploy SAM to automatically generate segmentation masks for all instances independently across different views.(b) For each segmented mask, we compute the instance center to construct the 2D Vote Map. (c)  By projecting 3D votes into pixel space through splatting and computing the voting loss together with the previous 2D voting map, each Gaussian primitive forming an instance is ensured to learn an offset vector pointing toward the instance center. For simplicity, we omit the rendering process and density control of other Gaussian parameters, as these are inherited from~\cite{kerbl20233d}.}
    \label{fig:pipeline}
    \vspace{-0.1in}
\end{figure*}

\section{Method}\label{sec:metho}

We now formally introduce VoteSplat, a 3D scene understanding framework for 3DGS representations, incorporating effective and distinctive voting mechanisms. We first employ SAM’s automatic mask generation module to produce masks for training views of the scene. The resulting multi-level mask information is then used to compute 2D ground-truth votes (detailed in Section.~\ref{sec:2d_votes}). Next, we embed offset vectors into 3DGS and train them to generate 3D votes (Section.~\ref{sec:3d_votes}). To improve depth consistency across different views, we introduce a depth regularization term, ensuring vote point aggregation during training. Finally, we establish a mapping between Gaussians and semantic information (Section.~\ref{sec:semantic_mapping}). The complete VoteSplat pipeline is illustrated in Figure~\ref{fig:pipeline}.

\subsection{Recap: 3D Gaussian Splatting}
Compared to implicit Neural Radiance Fields (NeRF)~\cite{mildenhall2021nerf}, 3D Gaussian Splatting (3DGS)~\cite{yu2024mip} constructs 3D scenes using explicit 3D Gaussian primitives. These primitives are represented as point clouds with associated attributes and are rendered through a tile-based differentiable rasterizer.

Given a training set of $K$ images $I = \{I_k\}_{k=1}^K$ with associated camera poses, and an image resolution of $H\times W$, the goal is to learn a set of  $N$  three-dimensional Gaussian, denoted as  $G = \{g_i\}_{i=1}^N$. Each Gaussian $g_i$ is characterized by quintet $g_i:=\{\mathbf{p_i}, \;\mathbf{s_i}, \;\mathbf{q_i}, \;o_i, \;\mathbf{c_i}\}$ of trainable parameters: the center position $\mathbf{p_i} \in \mathbb{R}^3$, the scaling factor $\mathbf{s_i} \in \mathbb{R}^3$, quaternion representing Gaussian's 3D covariance $\mathbf{q_i} \in \mathbb{R}^4$, opacity value $o_i \in \mathbb{R}$, and $\mathbf{c_i} \in [0,1]^3$ encodes RGB color using spherical harmonics coefficients.

After projecting 3D Gaussians onto the 2D image space under a given camera pose, 3DGS computes the pixel color  $\mathbf{C}$  using its differentiable rasterizer. The color is determined via $\alpha$-blending across $\mathcal{N}$ depth-ordered points overlapping the pixel, given by
\begin{equation}
\mathbf{C} = \sum_{i\in\mathcal{N}} \mathbf{c_i} \alpha_i T_i,
\end{equation}
where $\alpha_i$ is determined by evaluating the influence of each projected Gaussian based on its splatted 2D covariance $\Sigma$~\cite{yifan2019differentiable}, opacity $o_i$, and distance $\mathbf{d}$ to the pixel that reads
\begin{equation}
\alpha_i = o_i  \exp({-\frac{1}{2} \mathbf{d^T} \Sigma^{-1}\mathbf{d}}).    
\end{equation}
The transmittance $T_i:= \prod_{j=1}^{i-1} (1 - \alpha_j)$, represents the accumulated visibility of $i$-th Gaussian, accounting for occlusion from previously processed Gaussians in-depth order.

\subsection{2D Vote Construction}
\label{sec:2d_votes}
3D point clouds, such as those obtained from radar or Structure-from-Motion (SFM)~\cite{schonberger2016structure}, typically exhibit a concentration of points on the surface of spatial instances, with sparser distributions toward the instance centers.  This property persists during the densification process of 3DGS, wherein the generated point clouds remain concentrated on the instance surfaces. Consequently, directly clustering 3DGS point clouds can introduce boundary ambiguity between instances, leading to clustering errors when instance boundaries are in close proximity. \citet{shi2024language,choi2024click} embed additional feature vectors into the Gaussian Splatting framework to distinguish instances to avoid this. While effective, these methods often rely on computationally expensive contrastive learning. In contrast, 2D images inherently preserve structural information about instances and consistently capture object centers (provided the instance lies within the field of view). Exploiting this advantage, we propose a method to infer 3D instance centers from their corresponding 2D instance centers. We then detail the computation of the 2D centers (2D votes).

SAM effectively groups pixels belonging to the same instance and segment images into multiple object masks with well-defined boundaries. Following Lang-Splat, we utilize SAM to obtain precise hierarchical object masks. To improve the accuracy of 2D votes, we further filter out masks whose boundaries extend beyond the field of view (FoV).

For a given instance mask at level $l$, denoted as $M_l$, we compute the $x$-axis instance centroid $c_{x}^{l}$ by
\begin{equation}
c_{x}^{l} = \text{round}\left( \frac{ \sum_{y=0}^{H-1} \sum_{x=0}^{W-1} \; x \cdot M_{l}(x,y) }{\sum_{y=0}^{H-1} \sum_{x=0}^{W-1} \; M_{l}(x,y)} \right),
\end{equation}
recall $H$ and $W$ are the resolution of the image and $\mbox{round}(\cdot)$ is the rounding function and $y$-axis centroid $c_y^l$ is computed similarly. the 2D vote $\mathbf{V^{2d}_i}$ is thus defined as the doublet
\begin{equation}
\mathbf{V^{2d}_i}(x,y):= \{ c_x^l,\; c_y^l \}.
\end{equation}
Each pixel within the mask is then assigned to its corresponding 2D vote $\mathbf{V^{2d}_i}$, forming the ground-truth vote map, that serves as supervision for subsequent learning stages.

\subsection{3D Vote Construction} \label{sec:3d_votes}

{\bf 3D Vote Splatting.} In 3DGS, both the initial and densified point clouds predominantly lie on instance surfaces, making direct clustering-based instance separation challenging. To overcome this, we introduce an offset vector $\mathbf{\Delta p_i}\in\mathbb{R}^3$ for each Gaussian primitive, allowing the point cloud to vote toward instance centers. Consequently, the 3D vote reads
\begin{equation}
\mathbf{V^{3d}_i} = \mathbf{\Delta p_i}+\mathbf{p_i}.
\end{equation}
The additional vector attributes $\mathbf{V^{3d}_i}$ are typically optimized using the same $\alpha$-blending approach as color rendering, i.e.,
\begin{equation}
\mathbf{V^{3d}} = \sum_{i \in \mathcal{N}} \mathbf{V^{3d}_i} \alpha_i T_i.
\end{equation}
While generally effective, this approach faces two key challenges when applied to spatial constraints:
\begin{enumerate}
    \item[(i)] {\bf Unequal depth-weighting contribution:} In 3D space, all points forming an instance should contribute equally to its center vote. However, the traditional $\alpha$-weighting mechanism disproportionately reduces the influence of farther points, leading to biased voting.
    \item[(ii)] {\bf Incorrect occlusion Handling:}  In 3DGS, points behind an instance still affect the final color rendering but should not be involved in the voting process, as they do not belong to the instance.
\end{enumerate}
We introduce a distinct voting transmittance model to overcome this, applying uniform averaging in $\alpha$-blending as 
\begin{equation}
\mathbf{V^{3d}}:= \frac{1}{|\mathcal{M}|}\sum_{i \in \mathcal{M}} \mathbf{V^{3d}_i},
\end{equation}
where $\mathcal{M}$ represents set of the depth-ordered points under the voting transmittance $\hat{T}_i$ and $|\mathcal{M}|$ as its cardinality. Next, we project the blended 3D votes $\mathbf{V^{3d}}$ into screen space:
\begin{equation}
\mathbf{\tilde{V}^{2d}} =\mathbf{H}\mathbf{V^{3d}},
\end{equation}
where $\mathbf{H}$ is 4-by-4 transformation matrix from world space to screen space. Notably, blending is performed in 3D space before projection, ensuring stable voting. Otherwise, distant votes would experience significant fluctuations, making convergence toward the instance center less reliable.

The vote loss with respect to the precomputed 2D votes $\mathbf{V^{2d}}$ is defined as
\begin{equation}\label{eq:vote_loss}
\mathcal{L}_{vote}:= \frac{1}{|\mathcal{P}|}\sum_{i \in \mathcal{P}}|\mathbf{\tilde{V_i}^{2d}}-\mathbf{V_i^{2d}}|,
\end{equation}
where $\mathcal{P}$ denotes the set of pixels within the mask that contain 2D votes.\footnote{Throughout the paper, we adopted the notation $|\cdot|$ for $\ell_1$ norm.} Enforcing this loss ensures that offset vectors effectively guide Gaussians toward instance centers, while preserving the efficiency of the rasterization pipeline.

\begin{figure}[!t]
    \centering
    \includegraphics[width=\columnwidth]{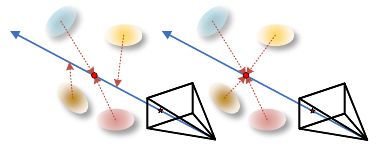}
    \caption{ Since projection results in-depth information loss, projected 3D votes may align correctly on the pixel plane but exhibit significant depth deviations, as shown in the left subfigure. The proposed depth regularization enforces spatial proximity along the depth dimension, keeping voting points concentrated in 3D space and enhancing overall voting consistency.
    }
    \label{fig:distort}
\end{figure}

{\bf \noindent Depth Regularization.} Projecting 3D votes to 2D votes inherently results in a loss of depth information. While training with multi-view images helps reduce depth uncertainty and encourages the point cloud to collectively vote toward the instance center, disturbances in the voting points may still persist, as shown in Figure~\ref{fig:distort}.
\label{Depth Regularization}

Inspired by depth distortion in 2DGS~\cite{huang20242d}, which promotes the concentration of Gaussian primitives, a parallel strategy is applied to 3D votes to preserve their depth alignment around the instance center. The initial depth distortion formulation in VoteSplat reads
\begin{equation}\label{eq:initial_distortion}
\mathcal{L}_{d} = \sum_{i,j}\; \omega_i\omega_j|z_i-z_j|.
\end{equation}
where $z$ is the $\mathbf{V^{3d}}$ depth in the camera coordinate system with weight $\omega_i:= \alpha_i T_i$.

Recall that the votes of Gaussian primitives need to be equally concentrated in-depth (av), the weight should not be reduced, and the formula is modified as 
\begin{equation}\label{eq:regularization_loss}
\mathcal{L}^*_{d} = \sum_{i,j}\;|z_i-z_j|.
\end{equation}
by eliminating the weights from \eqref{eq:initial_distortion}.

\textbf{Reconstruction-Voting-Depth (RVD) Loss:}  Lastly, the model optimizes the sparse point cloud, shifting it from its initial positions in the posed images toward a more concentrated target  with respect to the RVD loss, i.e.,
\begin{equation}
\mathcal{L} = \mathcal{L}_{c} +  \lambda_{vote}\mathcal{L}_{vote} +  \lambda_{L_{d}}\mathcal{L}^*_{d},
\end{equation}
where $\mathcal{L}_{c}(\mathcal{L}_{1},\mbox{D-SSIM})$ is the combination of RGB reconstruction loss $\mathcal{L}_{1}$~\cite[Eq. 7]{kerbl20233d} and D-SSIM term from~\cite[Eq. 7]{kerbl20233d}. The vote loss \eqref{eq:vote_loss} and depth regularization loss \eqref{eq:regularization_loss} penalized by their corresponding weights $\lambda_{vote}$ and $\lambda_{\mathcal L_{d}}$, respectively, both depends on instances' size in the scene.

\subsection{Semantic Construction}
\label{sec:semantic_mapping}
By constructing 3D votes, spatially separated voting points are obtained. To facilitate natural, open-vocabulary interactions, effectively associating 3D Gaussians with language features is essential. Thus, we propose an instance-level 3D-2D semantic association method based on voting points and instance IDs. The approach can be detailed as follows.
\begin{enumerate}
    \item[-] {\bf Background Filtering:} Point clouds with an offset vector $\mathbf{\Delta p} = \mathbf{0}$ are removed, as they are considered background and do not contribute to instance construction, receiving no gradient updates.
    \item[-] {\bf Clustering and Instance Dictionary Construction:} The remaining point clouds are clustered using HDBSCAN~\cite{mcinnes2017hdbscan} based on 3D votes, which performs density-based clustering while filtering out outliers. A dictionary is then built, where instance IDs serve as keys and the corresponding point cloud IDs as values.
    \item[-] {\bf Rendering Instance ID Maps:} The 3DGS rasterization pipeline is used to render the instance ID map. Combining this with the original RGB images, the pixel regions corresponding to each instance ID are identified.
    \item[-] {\bf Semantic Association with CLIP Features:} CLIP image features are extracted from the associated pixels, establishing a mapping between instance IDs and CLIP features, with multi-view feature integration incorporated for improved consistency.
\end{enumerate}

\begin{table*}[]
\centering
\small    
\scalebox{1}{
\begin{tabular}{l|cccc|c|cccc|c}
\toprule
\multirow{2}{*}{Methods}     & \multicolumn{5}{c}{mIoU $\uparrow$}    & \multicolumn{5}{c}{mAcc.  $\uparrow$} \\
& figurines    & teatime    & ramen     & waldo\_kitchen    & Mean   & figurines   & teatime   & ramen   & waldo\_kitchen  & Mean\\
\midrule
LangSplat      & 10.16      & 11.38       & 7.92     & 9.18      & 9.66     & 8.93       & 20.34    & 11.27    & 9.09    & 12.41\\
OpenGaussian   & 60.11      & 65.80       & 31.01    & 22.70     & 44.90    & 82.14      & 79.66    & 42.25    & 31.92   & 58.99\\
VoteSplat     & \textbf{68.62} & \textbf{66.71} &\textbf{39.24} & \textbf{25.84} & \textbf{50.10} & \textbf{85.71} & \textbf{88.14} & \textbf{61.97} & \textbf{33.68} & \textbf{67.38}\\
\bottomrule
\end{tabular}}
    \caption{Performance of semantic segmentation on the LeRF dataset compared to LangSplat and OpenGaussian based on text query. Accuracy is measured by mAcc@0.25.}
    \label{table:sem1}
\end{table*}

\begin{table}[t]
\centering
\scalebox{1}{
\begin{tabular}{l|cccc}
\toprule
Methods & snacks  & figurines & teatime & ramen \\
\midrule
LS(Level1) & $\sim$ 67 & $\sim$ 116   & $\sim$ 87 & $\sim$ 56 \\
OG   & $\sim$ 114 & $\sim$117   & $\sim$ 104 & $\sim$55 \\
GG   & -       & $\sim$ 150    &$\sim$122 & $\sim$ 90 \\
VS   &   $\sim$ \textbf{57}     &   $\sim$\textbf{ 54}       &   $\sim$\textbf{ 43}      & $\sim$\textbf{ 53}   \\
\bottomrule
\end{tabular}}
    \caption{The training time (in minutes) of LangSplat (LS), OpenGaussian (OG), GaussianGrouping (GG), and VoteSplat (VS).\protect\footnotemark}
    \label{table:time}
\end{table}

\begin{figure*}[htb!]
    \centering
    \includegraphics[width=1\textwidth]{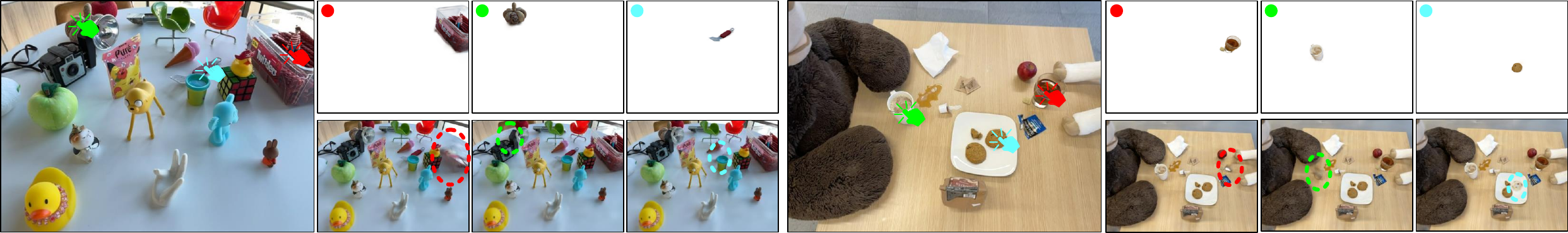}
    \caption{Click-based 3D object selection and scene editing results. VoteSplat enables complete 3D object selection without issues of incompleteness or redundancy. Moreover, after removing the selected Gaussian primitives, the scene can be effectively edited.}
    \label{fig:click}
\end{figure*}

\section{Experiments}\label{sec:exp}

\subsection{Open-Vocabulary Object Selection in 3D Space}\label{OV}

\paragraph{Experimental Setup.} 

\textbf{(i) Objective:} Given an open-vocabulary text query, CLIP extracts textual features, and cosine similarity is computed with each instance ID’s language features. The most relevant instances are selected, and their Gaussian primitives are rendered into multi-view images via the 3DGS rasterization pipeline;
\textbf{(ii) Baseline:} Our method is compared against LangSplat, Gaussian Grouping, and OpenGaussian. VoteSplat follows the method described in Section~\ref{sec:semantic_mapping} to associate each instance with a 512-dimensional CLIP feature and select the corresponding Gaussian primitives for rendering. For LangSplat and OpenGaussian, we adhere to their respective procedures. In LangSplat, the 512-dimensional CLIP feature is reconstructed from the low-dimensional language feature of each Gaussian. In OpenGaussian, cosine similarity is computed to select the corresponding Gaussian primitives for rendering. Since Gaussian Grouping does not inherently support semantic queries on Gaussian primitives and is limited to instance segmentation, it is excluded from semantic query comparisons;
\textbf{(iii) Dataset and Metrics:} Experiments are conducted on 3D-OVS\cite{liu2023weakly} and Lerf-OVS\cite{kerr2023lerf}, with all datasets annotated by LangSplat. Performance is evaluated using average IoU and segmentation accuracy, measuring the alignment between rendered images (from selected 3D Gaussian points) and ground-truth object masks. Additionally, we report training time across different methods and provide feature visualizations of the point clouds.

\textbf{Result.}
Table~\ref{table:sem1} shows VoteSplat outperforms other methods in both mIoU and mAcc. LangSplat, with weaker 3D understanding, struggles to accurately associate 3D Gaussian points with query text. This suggests that 2D image semantics derived from $\alpha$-blending fail to effectively capture 3D semantics encoded by Gaussian primitives. Table~\ref{table:time} \footnotetext{The running time is measured in minutes, with seconds omitted, denoted by the symbol $\sim$.} reports the training time for each model, evaluated over 60,000 iterations on NVIDIA RTX 3090 GPU. VoteSplat achieves the shortest training time among all methods, benefiting from its efficient voting mechanism. In contrast, OpenGaussian’s intra-/inter-class contrastive learning significantly increases training time. Gaussian Grouping~\cite{ye2024gaussian} relies on KNN for feature consistency, leading to high computational and memory complexity during training.

\begin{figure*}[htb!]
    \centering
\includegraphics[width=0.97\textwidth]{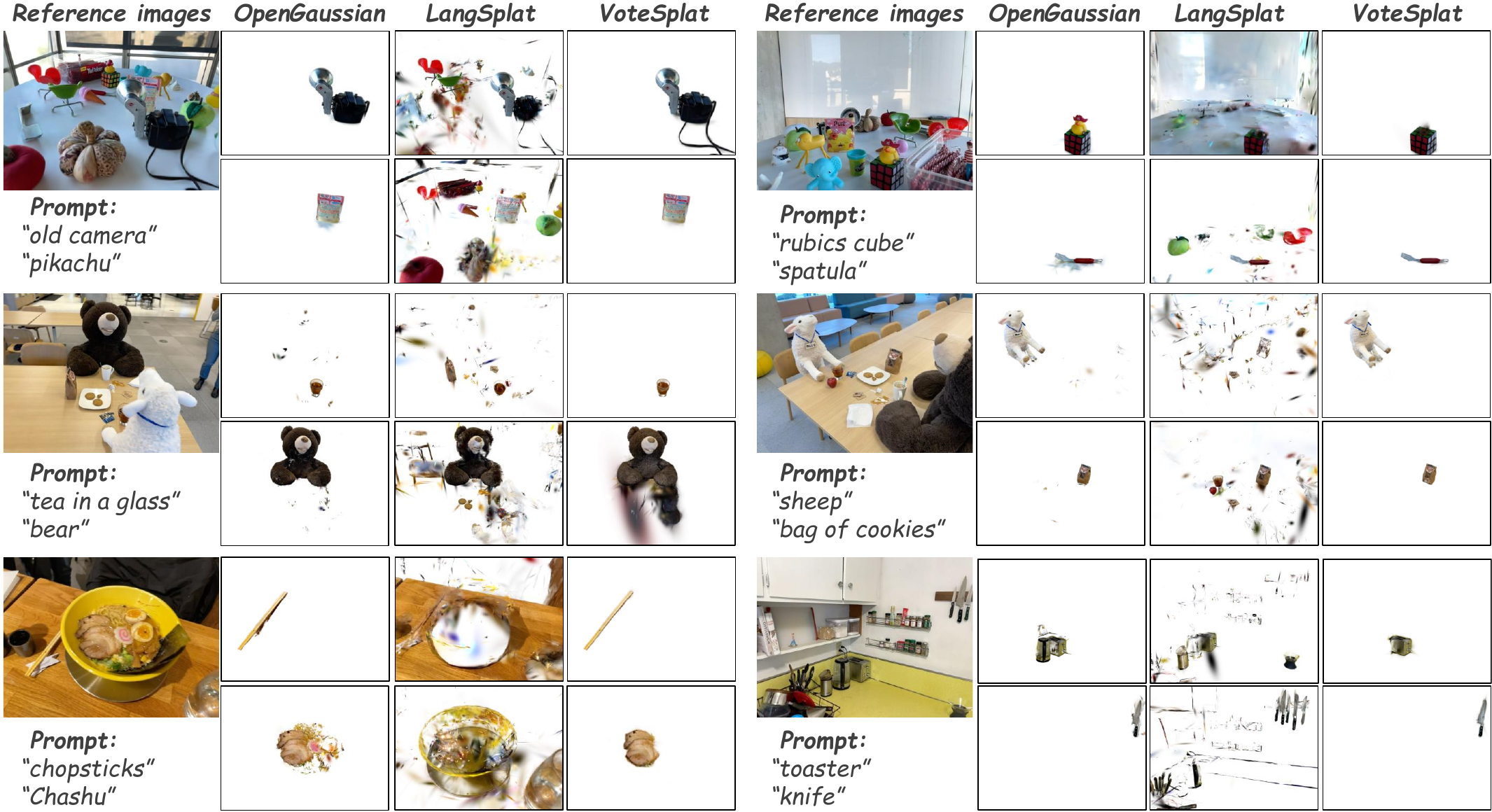}
    \caption{Open-vocabulary 3D object selection on the LERF dataset. VoteSplat outperforms LangSplat and OpenGaussian in accurately identifying 3D objects corresponding to text queries.}
    \label{fig:sem1}
\end{figure*}

Qualitative results are presented in~Figure~\ref{fig:sem1}. Given a text query, VoteSplat selects relevant Gaussian points and renders them into multi-view images. Due to the ambiguity of 3D point features, LangSplat struggles to accurately recognize target objects, while OpenGaussian fails to capture fine-grained details as effectively as VoteSplat. For example, with the prompt “old camera,” VoteSplat successfully clusters finer details, such as the rope. Additionally, VoteSplat outperforms other methods in rendering occluded objects. In the teatime scene, it reconstructs the bear’s lower body, despite being partially obscured by the table. Moreover, VoteSplat generates images with fewer noise artifacts, enhancing overall rendering quality.

\begin{figure*}[htb!]
    \centering
    \includegraphics[width=0.9\textwidth]{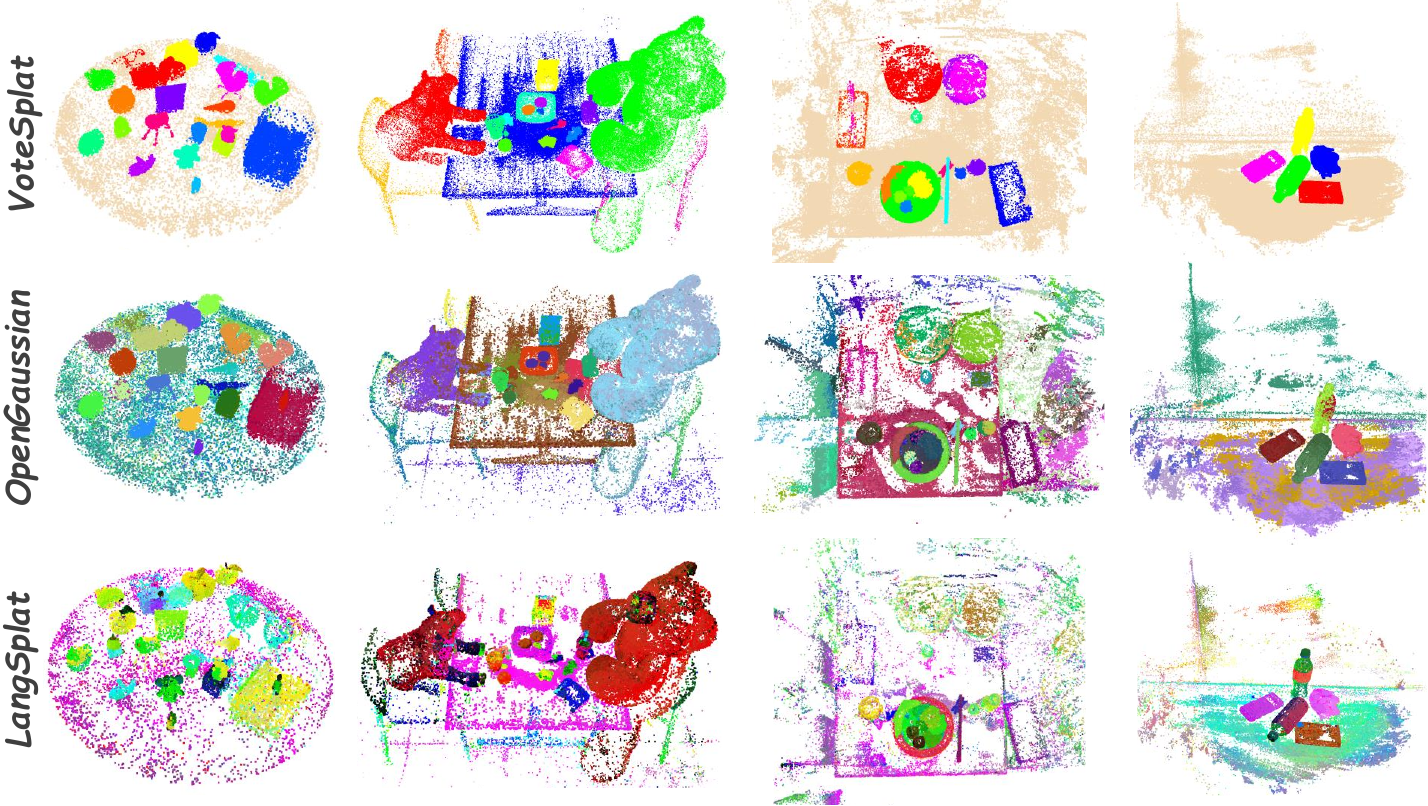}
    \caption{Comparison of point cloud feature visualizations. From left to right, the scenes correspond to \textit{figurines, teatime, ramen, and snacks}. The first three scenes are from LeRF, and the last scene is from 3D-OVS. Our proposed method, VoteSplat, demonstrates superior performance in terms of feature granularity and accuracy.}
    \label{fig:point}
    \vspace{-0.1in}
\end{figure*}

Figure~\ref{fig:point} visualizes the point cloud features, where VoteSplat assigns distinct colors to instance categories for clarity. In OpenGaussian, colors are derived by applying PCA to reduce feature dimensions to three, and then mapping them to RGB. In contrast, LangSplat directly uses its three-dimensional features as point cloud colors. 
The feature visualization for Gaussian Grouping is provided in the supplementary material. The well-segmented instances in VoteSplat demonstrate its superior performance.

\subsection{Click-based 3D Object Selection and Editing}

Given an image from any viewpoint, clicking on a 2D pixel selects the corresponding 3D Gaussian points. The instance ID associated with the selected Gaussian primitive is then retrieved, enabling click-based object selection. Additionally, removing the entire instance allows for scene editing effects. Figure~\ref{fig:click} demonstrates click-based object selection and scene editing on the LERF dataset. The left image highlights instance segmentation under occlusion, while the right image focuses on small object selection.

\subsection{Hierarchical Segmentation}
\begin{figure*}[htb!]
    \centering
    \includegraphics[width=1\textwidth]{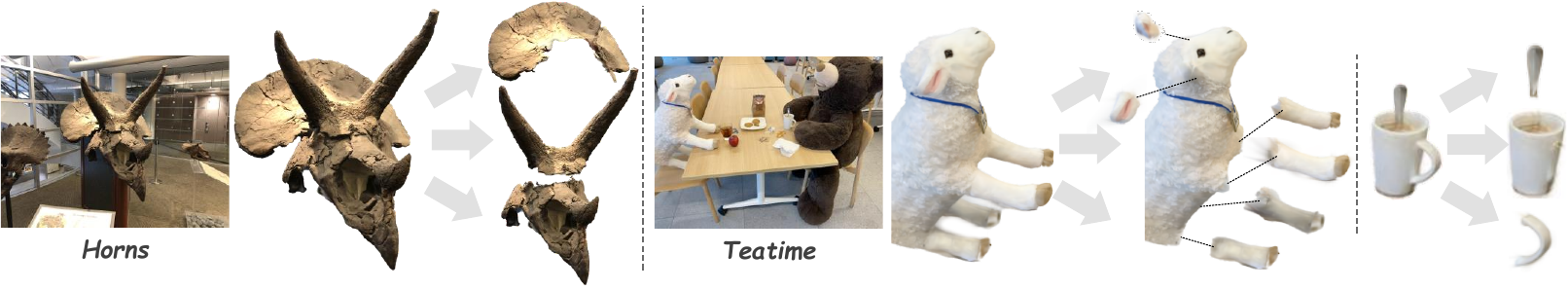}
    \caption{Using SAM, objects are divided into multiple parts, each assigned a 2D vote. After training, VoteSplat generates a corresponding 3D vote for each component. The rendering results of these components are then visualized.}
    \label{fig:example}
    \vspace{-0.1in}
\end{figure*}

In some cases, instance segmentation is sufficient, but certain applications require finer segmentation, such as part segmentation for more precise analysis. VoteSplat supports this functionality by utilizing multi-level masks from SAM to compute hierarchical 2D votes. These votes generate layered 2D voting maps, which then supervise 3D votes, enabling finer-grained segmentation.
As shown in Figure~\ref{fig:example}, hierarchical 3D votes and rendering results are presented on the LLFF~\cite{mildenhall2021nerf} and LeRF datasets.

\begin{figure*}[htb!]
    \centering
    \includegraphics[width=1\textwidth]{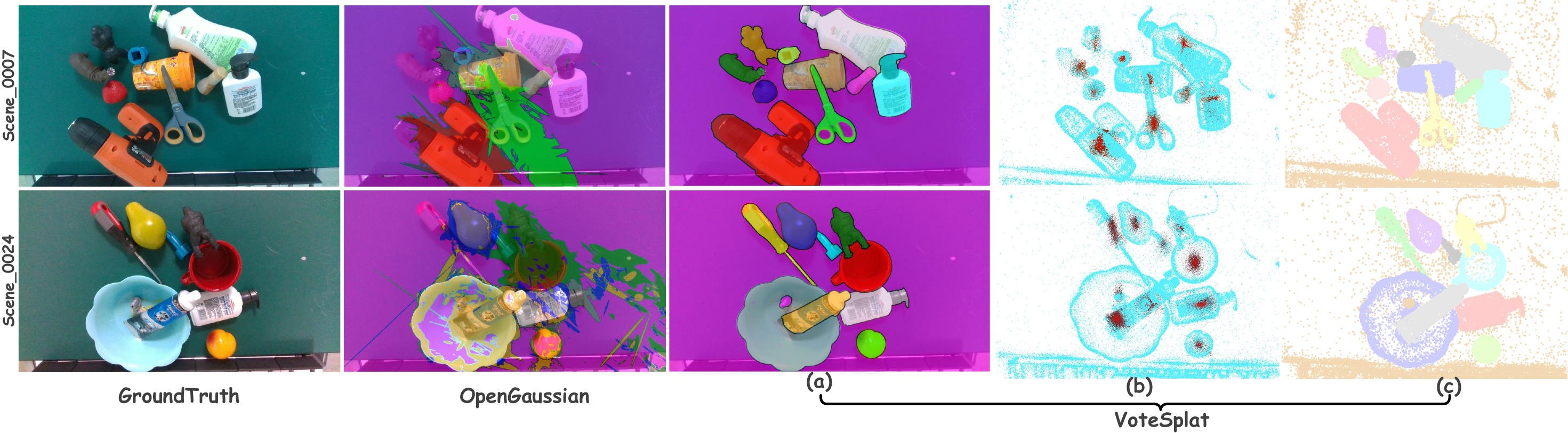}
    \caption{Instance segmentation in complex scenes. Compared with Open-Gaussian, VoteSplat can handle more complex scenarios, such as instance overlap and contain each other.}
    \label{fig:complex}
     \vspace{-0.15in}
\end{figure*}

\begin{figure}[t]
    \centering
    \includegraphics[width=\columnwidth]{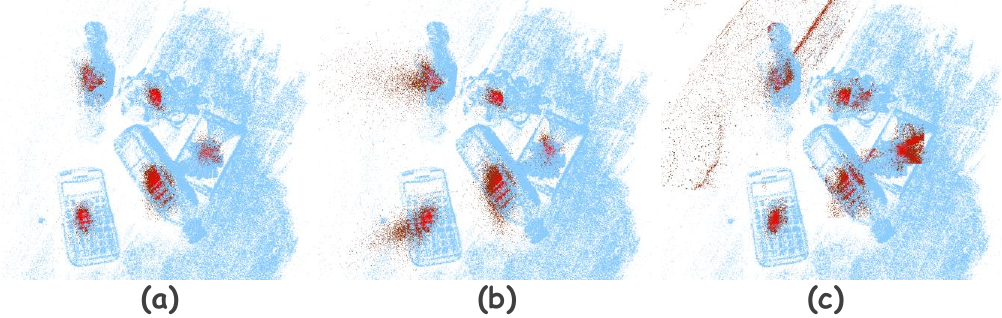}
    \caption{Comparison of ablation experiments. (a) shows 3D votes of VoteSplat are precisely located at the instance center. (b) reflects the effects of \textit{w/o} $\mathcal{L}_{d}$ and the consequence of projecting first and then accumulating $\mathbf{V^{3d}_i}$. (c) illustrates the impact of using transmittance $T$.}
    \label{fig:ablation}
    \vspace{-0.15in}
\end{figure}

\subsection{Instance Segmentation in Complex Scenes}
To evaluate VoteSplat in complex scenarios, experiments is conducted on GraspNet dataset~\cite{fang2020graspnet}, where instances are overlapping, adjacent, and contained. Despite instances' proximity, VoteSplat successfully segments them. As shown in Figure~\ref{fig:complex}(b), 3D votes remain well-separated, demonstrating the method’s effectiveness.

\subsection{Ablation Study}
As discussed in Section~\ref{Depth Regularization}, projection inherently leads to depth information loss. While the projected points may align with 2D votes on the pixel plane, they can exhibit significant depth discrepancies. This issue is more pronounced in forward-facing data compared to 360-degree captures, as the absence of side-view images prevents voting points from consistently converging toward the instance center. Figure~\ref{fig:ablation} illustrates the impact of $\mathcal{L}_{d}$ on the 3D-OVS dataset. Without $\mathcal{L}_{d}$, the point cloud appears dispersed along the depth dimension, whereas with $\mathcal{L}_{d}$, the points are more tightly clustered.
Similarly, applying projection before accumulation results in 3D vote dispersion. Additionally, when using transmittance $T$, background points participate in voting, introducing clustering disturbances, as in Figure~\ref{fig:ablation}(c). 
The quantitative results are reported in Table~\ref{tab:ablation}. It also demonstrates that each component of our method contributes to segmentation performance. In particular, allowing background Gaussians to participate in the voting process leads to a significant degradation in segmentation accuracy, highlighting the importance of explicitly filtering them out.
\begin{table}[t]
\centering
\begin{tabular}{lcccc}
\toprule
Case & Eq. 7 & L\_d & mIoU(\%)$\uparrow$ & mAcc.$\uparrow$ \\
\midrule
\#1  &  \checkmark     &      & 76.04 & 0.88 \\
\#2  &       &   \checkmark   & 58.72 & 0.72 \\
\#3  &  \checkmark     &   \checkmark   & \textbf{85.66} & \textbf{0.96} \\
\bottomrule
\end{tabular}
\caption{Quantitative ablations result on the snacks scene of the 3D-OVS dataset.}
\label{tab:ablation}
\end{table}
\section{Concluding Remarks}\label{sec:con}
We introduce VoteSplat, a novel 3D scene understanding method that integrates Hough voting with 3D Gaussian Splatting (3DGS). Utilizing SAM for image instance segmentation, we generate 2D vote maps to supervise 3D votes, which are computed through embedded spatial offset vectors. To further refine clustering, we introduce depth distortion, constraining spatial offsets along the depth dimension, ensuring Gaussian primitives are well-clustered in 3D space. Additionally, by projecting instance IDs, VoteSplat establishes a precise correspondence between Gaussian primitives and 2D image semantics, effectively resolving semantic ambiguities. Experimental results confirm its effectiveness across various tasks. 
The voting mechanism encounters challenges when instances have projected sizes that significantly exceed the field of view (FoV), leading to inaccuracies in 2D votes. Additionally, it may also struggle with instances enclosed in concave containers, where spatial proximity makes precise separation difficult.

\section*{Acknowledgement}
The authors sincerely thank the anonymous reviewers for their valuable comments and suggestions. This work was supported by grants from the Natural Science Foundation of Shanxi Province (2024JCJCQN-66),  Science and Technology Commission of Shanghai Municipality (NO.24511106900), and is partially supported by the National Natural Science Foundation of China under grant Nos. 62072358 and 62072352. 
{
    \small
    \bibliographystyle{template/ieeenat_fullname}
    \bibliography{references}

\begin{thebibliography}{42}
\providecommand{\natexlab}[1]{#1}
\providecommand{\url}[1]{\texttt{#1}}
\expandafter\ifx\csname urlstyle\endcsname\relax
  \providecommand{\doi}[1]{doi: #1}\else
  \providecommand{\doi}{doi: \begingroup \urlstyle{rm}\Url}\fi

\bibitem[Ballard(1981)]{ballard1981generalizing}
Dana~H. Ballard.
\newblock Generalizing the {H}ough transform to detect arbitrary shapes.
\newblock \emph{Pattern recognition}, 13\penalty0 (2):\penalty0 111--122, 1981.

\bibitem[Borrmann et~al.(2011)Borrmann, Elseberg, Lingemann, and N{\"u}chter]{borrmann20113d}
Dorit Borrmann, Jan Elseberg, Kai Lingemann, and Andreas N{\"u}chter.
\newblock The 3{D} {H}ough transform for plane detection in point clouds: A review and a new accumulator design.
\newblock \emph{3{D} Research}, 2\penalty0 (2):\penalty0 1--13, 2011.

\bibitem[Caron et~al.(2021)Caron, Touvron, Misra, J{\'e}gou, Mairal, Bojanowski, and Joulin]{caron2021emerging}
Mathilde Caron, Hugo Touvron, Ishan Misra, Herv{\'e} J{\'e}gou, Julien Mairal, Piotr Bojanowski, and Armand Joulin.
\newblock Emerging properties in self-supervised vision transformers.
\newblock In \emph{Proceedings of the IEEE/CVF international conference on computer vision}, pages 9650--9660, 2021.

\bibitem[Chen et~al.(2024)Chen, Li, Ye, Wang, Xie, Zhai, Wang, Liu, Bao, and Zhang]{chen2024pgsr}
Danpeng Chen, Hai Li, Weicai Ye, Yifan Wang, Weijian Xie, Shangjin Zhai, Nan Wang, Haomin Liu, Hujun Bao, and Guofeng Zhang.
\newblock P{GSR}: Planar-based gaussian splatting for efficient and high-fidelity surface reconstruction.
\newblock \emph{arXiv:2406.06521}, 2024.

\bibitem[Choi et~al.(2024)Choi, Song, Kim, Kim, and Do]{choi2024click}
Seokhun Choi, Hyeonseop Song, Jaechul Kim, Taehyeong Kim, and Hoseok Do.
\newblock Click-gaussian: Interactive segmentation to any 3{D} gaussians.
\newblock In \emph{European Conference on Computer Vision}, pages 289--305. Springer, 2024.

\bibitem[Dong et~al.(2024)Dong, Li, Huang, Bian, Liu, Bao, Cui, Li, and Zhang]{dong2024global}
Yitong Dong, Yijin Li, Zhaoyang Huang, Weikang Bian, Jingbo Liu, Hujun Bao, Zhaopeng Cui, Hongsheng Li, and Guofeng Zhang.
\newblock A global depth-range-free multi-view stereo transformer network with pose embedding.
\newblock \emph{arXiv preprint arXiv:2411.01893}, 2024.

\bibitem[Fang et~al.(2020)Fang, Wang, Gou, and Lu]{fang2020graspnet}
Hao-Shu Fang, Chenxi Wang, Minghao Gou, and Cewu Lu.
\newblock Graspnet-1billion: A large-scale benchmark for general object grasping.
\newblock In \emph{Proceedings of the IEEE/CVF conference on computer vision and pattern recognition}, pages 11444--11453, 2020.

\bibitem[Hough(1959)]{hough1959machine}
Paul~V.C. Hough.
\newblock Machine analysis of bubble chamber pictures.
\newblock In \emph{International Conference on High Energy Accelerators and Instrumentation, CERN, 1959}, pages 554--556, 1959.

\bibitem[Huang et~al.(2024)Huang, Yu, Chen, Geiger, and Gao]{huang20242d}
Binbin Huang, Zehao Yu, Anpei Chen, Andreas Geiger, and Shenghua Gao.
\newblock 2{D} gaussian splatting for geometrically accurate radiance fields.
\newblock In \emph{ACM SIGGRAPH 2024 conference papers}, pages 1--11, 2024.

\bibitem[Kerbl et~al.(2023)Kerbl, Kopanas, Leimk{\"u}hler, and Drettakis]{kerbl20233d}
Bernhard Kerbl, Georgios Kopanas, Thomas Leimk{\"u}hler, and George Drettakis.
\newblock 3{D} gaussian splatting for real-time radiance field rendering.
\newblock \emph{ACM Trans. Graph.}, 42\penalty0 (4):\penalty0 139--1, 2023.

\bibitem[Kerr et~al.(2023)Kerr, Kim, Goldberg, Kanazawa, and Tancik]{kerr2023lerf}
Justin Kerr, Chung~Min Kim, Ken Goldberg, Angjoo Kanazawa, and Matthew Tancik.
\newblock Lerf: Language embedded radiance fields.
\newblock In \emph{Proceedings of the IEEE/CVF International Conference on Computer Vision}, pages 19729--19739, 2023.

\bibitem[Kirillov et~al.(2023)Kirillov, Mintun, Ravi, Mao, Rolland, Gustafson, Xiao, Whitehead, Berg, Lo, Doll\'{a}r, and Girshick]{kirillov2023segment}
Alexander Kirillov, Eric Mintun, Nikhila Ravi, Hanzi Mao, Chloe Rolland, Laura Gustafson, Tete Xiao, Spencer Whitehead, Alexander~C. Berg, Wan-Yen Lo, Piotr Doll\'{a}r, and Ross Girshick.
\newblock Segment anything.
\newblock In \emph{Proceedings of the IEEE/CVF international conference on computer vision}, pages 4015--4026, 2023.

\bibitem[Knopp et~al.(2010)Knopp, Prasad, and Van~Gool]{knopp2010orientation}
Jan Knopp, Mukta Prasad, and Luc Van~Gool.
\newblock Orientation invariant 3{D} object classification using {H}ough transform based methods.
\newblock In \emph{Proceedings of the ACM workshop on 3{D} object retrieval}, pages 15--20, 2010.

\bibitem[Knopp et~al.(2011)Knopp, Prasad, and Van~Gool]{knopp2011scene}
Jan Knopp, Mukta Prasad, and Luc Van~Gool.
\newblock Scene cut: Class-specific object detection and segmentation in 3{D} scenes.
\newblock In \emph{2011 International Conference on 3{D} Imaging, Modeling, Processing, Visualization and Transmission}, pages 180--187. IEEE, 2011.

\bibitem[Leibe et~al.(2008)Leibe, Leonardis, and Schiele]{leibe2008robust}
Bastian Leibe, Ale{\v{s}} Leonardis, and Bernt Schiele.
\newblock Robust object detection with interleaved categorization and segmentation.
\newblock \emph{International journal of computer vision}, 77:\penalty0 259--289, 2008.

\bibitem[Li et~al.(2024)Li, Wu, Meng, Gao, Zhang, Wang, and Zhang]{li2024instancegaussian}
Haijie Li, Yanmin Wu, Jiarui Meng, Qiankun Gao, Zhiyao Zhang, Ronggang Wang, and Jian Zhang.
\newblock Instancegaussian: Appearance-semantic joint gaussian representation for 3{D} instance-level perception.
\newblock \emph{arXiv:2411.19235}, 2024.

\bibitem[Liu et~al.(2023)Liu, Zhan, Zhang, Xu, Yu, El~Saddik, Theobalt, Xing, and Lu]{liu2023weakly}
Kunhao Liu, Fangneng Zhan, Jiahui Zhang, Muyu Xu, Yingchen Yu, Abdulmotaleb El~Saddik, Christian Theobalt, Eric Xing, and Shijian Lu.
\newblock Weakly supervised 3{D} open-vocabulary segmentation.
\newblock \emph{Advances in Neural Information Processing Systems}, 36:\penalty0 53433--53456, 2023.

\bibitem[Lu et~al.(2024)Lu, Yu, Xu, Xiangli, Wang, Lin, and Dai]{lu2024scaffold}
Tao Lu, Mulin Yu, Linning Xu, Yuanbo Xiangli, Limin Wang, Dahua Lin, and Bo Dai.
\newblock Scaffold-{GS}: Structured 3{D} gaussians for view-adaptive rendering.
\newblock In \emph{Proceedings of the IEEE/CVF Conference on Computer Vision and Pattern Recognition}, pages 20654--20664, 2024.

\bibitem[McInnes et~al.(2017)McInnes, Healy, and Astels]{mcinnes2017hdbscan}
Leland McInnes, John Healy, and Steve Astels.
\newblock hdbscan: Hierarchical density based clustering.
\newblock \emph{J. Open Source Softw.}, 2\penalty0 (11):\penalty0 205, 2017.

\bibitem[Mildenhall et~al.(2021)Mildenhall, Srinivasan, Tancik, Barron, Ramamoorthi, and Ng]{mildenhall2021nerf}
Ben Mildenhall, Pratul~P. Srinivasan, Matthew Tancik, Jonathan~T. Barron, Ravi Ramamoorthi, and Ren Ng.
\newblock Nerf: Representing scenes as neural radiance fields for view synthesis.
\newblock \emph{Communications of the ACM}, 65\penalty0 (1):\penalty0 99--106, 2021.

\bibitem[Qi et~al.(2017{\natexlab{a}})Qi, Su, Mo, and Guibas]{qi2017pointnet}
Charles~R. Qi, Hao Su, Kaichun Mo, and Leonidas~J. Guibas.
\newblock Pointnet: Deep learning on point sets for 3{D} classification and segmentation.
\newblock In \emph{Proceedings of the IEEE conference on computer vision and pattern recognition}, pages 652--660, 2017{\natexlab{a}}.

\bibitem[Qi et~al.(2017{\natexlab{b}})Qi, Yi, Su, and Guibas]{qi2017pointnet++}
Charles~Ruizhongtai Qi, Li Yi, Hao Su, and Leonidas~J. Guibas.
\newblock Pointnet++: Deep hierarchical feature learning on point sets in a metric space.
\newblock \emph{Advances in neural information processing systems}, 30, 2017{\natexlab{b}}.

\bibitem[Qi et~al.(2019)Qi, Litany, He, and Guibas]{qi2019deep}
Charles~R. Qi, Or Litany, Kaiming He, and Leonidas~J. Guibas.
\newblock Deep {H}ough voting for 3{D} object detection in point clouds.
\newblock In \emph{proceedings of the IEEE/CVF International Conference on Computer Vision}, pages 9277--9286, 2019.

\bibitem[Qin et~al.(2024)Qin, Li, Zhou, Wang, and Pfister]{qin2024langsplat}
Minghan Qin, Wanhua Li, Jiawei Zhou, Haoqian Wang, and Hanspeter Pfister.
\newblock Langsplat: 3{D} language gaussian splatting.
\newblock In \emph{Proceedings of the IEEE/CVF Conference on Computer Vision and Pattern Recognition}, pages 20051--20060, 2024.

\bibitem[Qiu et~al.(2024)Qiu, Liu, Su, and Lin]{qiu2024gls}
Jiaxiong Qiu, Liu Liu, Zhizhong Su, and Tianwei Lin.
\newblock G{LS}: Geometry-aware 3{D} language gaussian splatting.
\newblock \emph{arXiv preprint arXiv:2411.18066}, 2024.

\bibitem[Redmon et~al.(2016)Redmon, Divvala, Girshick, and Farhadi]{redmon2016you}
Joseph Redmon, Santosh Divvala, Ross Girshick, and Ali Farhadi.
\newblock You only look once: Unified, real-time object detection.
\newblock In \emph{Proceedings of the IEEE conference on computer vision and pattern recognition}, pages 779--788, 2016.

\bibitem[Ren et~al.(2016)Ren, He, Girshick, and Sun]{ren2016faster}
Shaoqing Ren, Kaiming He, Ross Girshick, and Jian Sun.
\newblock Faster {R}-{CNN}: Towards real-time object detection with region proposal networks.
\newblock \emph{IEEE transactions on pattern analysis and machine intelligence}, 39\penalty0 (6):\penalty0 1137--1149, 2016.

\bibitem[Schonberger and Frahm(2016)]{schonberger2016structure}
Johannes~L. Schonberger and Jan-Michael Frahm.
\newblock Structure-from-motion revisited.
\newblock In \emph{Proceedings of the IEEE conference on computer vision and pattern recognition}, pages 4104--4113, 2016.

\bibitem[Shi et~al.(2024)Shi, Wang, Duan, and Guan]{shi2024language}
Jin-Chuan Shi, Miao Wang, Hao-Bin Duan, and Shao-Hua Guan.
\newblock Language embedded 3{D} gaussians for open-vocabulary scene understanding.
\newblock In \emph{Proceedings of the IEEE/CVF Conference on Computer Vision and Pattern Recognition}, pages 5333--5343, 2024.

\bibitem[Shorinwa et~al.(2024)Shorinwa, Sun, and Schwager]{shorinwa2024fast}
Ola Shorinwa, Jiankai Sun, and Mac Schwager.
\newblock Fast-{S}plat: Fast, ambiguity-free semantics transfer in gaussian splatting.
\newblock \emph{arXiv:2411.13753}, 2024.

\bibitem[Siddiqui et~al.(2023)Siddiqui, Porzi, Bul{\'o}, M{\"u}ller, Nie{\ss}ner, Dai, and Kontschieder]{siddiqui2023panoptic}
Yawar Siddiqui, Lorenzo Porzi, Samuel~Rota Bul{\'o}, Norman M{\"u}ller, Matthias Nie{\ss}ner, Angela Dai, and Peter Kontschieder.
\newblock Panoptic lifting for 3{D} scene understanding with neural fields.
\newblock In \emph{Proceedings of the IEEE/CVF Conference on Computer Vision and Pattern Recognition}, pages 9043--9052, 2023.

\bibitem[Sun et~al.(2010)Sun, Bradski, Xu, and Savarese]{sun2010depth}
Min Sun, Gary Bradski, Bing-Xin Xu, and Silvio Savarese.
\newblock Depth-encoded {H}ough voting for joint object detection and shape recovery.
\newblock In \emph{European Conference on Computer Vision}, pages 658--671. Springer, 2010.

\bibitem[Velizhev et~al.(2012)Velizhev, Shapovalov, and Schindler]{velizhev2012implicit}
Alexander Velizhev, Roman Shapovalov, and Konrad Schindler.
\newblock Implicit shape models for object detection in 3{D} point clouds.
\newblock \emph{ISPRS Annals of the Photogrammetry, Remote Sensing and Spatial Information Sciences}, 1:\penalty0 179--184, 2012.

\bibitem[Woodford et~al.(2014)Woodford, Pham, Maki, Perbet, and Stenger]{woodford2014demisting}
Oliver~J. Woodford, Minh-Tri Pham, Atsuto Maki, Frank Perbet, and Bj{\"o}rn Stenger.
\newblock Demisting the {H}ough transform for 3{D} shape recognition and registration.
\newblock \emph{International Journal of Computer Vision}, 106:\penalty0 332--341, 2014.

\bibitem[Wu et~al.(2024)Wu, Meng, Li, Wu, Shi, Cheng, Zhao, Feng, Ding, Wang, and Zhang]{wu2024opengaussian}
Yanmin Wu, Jiarui Meng, Haijie Li, Chenming Wu, Yahao Shi, Xinhua Cheng, Chen Zhao, Haocheng Feng, Errui Ding, Jingdong Wang, and Jian Zhang.
\newblock Opengaussian: Towards point-level 3{D} gaussian-based open vocabulary understanding.
\newblock \emph{arXiv:2406.02058}, 2024.

\bibitem[Ye et~al.(2024)Ye, Danelljan, Yu, and Ke]{ye2024gaussian}
Mingqiao Ye, Martin Danelljan, Fisher Yu, and Lei Ke.
\newblock Gaussian grouping: Segment and edit anything in 3{D} scenes.
\newblock In \emph{European Conference on Computer Vision}, pages 162--179. Springer, 2024.

\bibitem[Yi et~al.(2023)Yi, Fang, Wu, Xie, Zhang, Liu, Tian, and Wang]{yi2023gaussiandreamer}
Taoran Yi, Jiemin Fang, Guanjun Wu, Lingxi Xie, Xiaopeng Zhang, Wenyu Liu, Qi Tian, and Xinggang Wang.
\newblock Gaussiandreamer: {F}ast generation from text to 3{D} gaussian splatting with point cloud priors.
\newblock \emph{arXiv:2310.08529}, 2023.

\bibitem[Yifan et~al.(2019)Yifan, Serena, Wu, {\"O}ztireli, and Sorkine-Hornung]{yifan2019differentiable}
Wang Yifan, Felice Serena, Shihao Wu, Cengiz {\"O}ztireli, and Olga Sorkine-Hornung.
\newblock Differentiable surface splatting for point-based geometry processing.
\newblock \emph{ACM Transactions On Graphics (TOG)}, 38\penalty0 (6):\penalty0 1--14, 2019.

\bibitem[Yu et~al.(2024{\natexlab{a}})Yu, Chen, Huang, Sattler, and Geiger]{yu2024mip}
Zehao Yu, Anpei Chen, Binbin Huang, Torsten Sattler, and Andreas Geiger.
\newblock Mip-splatting: Alias-free 3{D} gaussian splatting.
\newblock In \emph{Proceedings of the IEEE/CVF Conference on Computer Vision and Pattern Recognition}, pages 19447--19456, 2024{\natexlab{a}}.

\bibitem[Yu et~al.(2024{\natexlab{b}})Yu, Sattler, and Geiger]{yu2024gaussian}
Zehao Yu, Torsten Sattler, and Andreas Geiger.
\newblock Gaussian opacity fields: Efficient and compact surface reconstruction in unbounded scenes.
\newblock \emph{arXiv:2404.10772}, 2024{\natexlab{b}}.

\bibitem[Zhai et~al.(2025)Zhai, Li, Li, Pan, He, and Zhang]{zhai_cvpr25_panogs}
Hongjia Zhai, Hai Li, Zhenzhe Li, Xiaokun Pan, Yijia He, and Guofeng Zhang.
\newblock Panogs: Gaussian-based panoptic segmentation for 3d open vocabulary scene understanding.
\newblock In \emph{Proceedings of the Computer Vision and Pattern Recognition Conference (CVPR)}, pages 14114--14124, 2025.

\bibitem[Zhi et~al.(2021)Zhi, Laidlow, Leutenegger, and Davison]{zhi2021place}
Shuaifeng Zhi, Tristan Laidlow, Stefan Leutenegger, and Andrew~J. Davison.
\newblock In-place scene labelling and understanding with implicit scene representation.
\newblock In \emph{Proceedings of the IEEE/CVF International Conference on Computer Vision}, pages 15838--15847, 2021.

\end{thebibliography}
}

\end{document}